\documentclass[11pt]{article}

\usepackage[utf8]{inputenc}
\usepackage[T1]{fontenc}
\usepackage{amsmath,amssymb}
\usepackage{graphicx}
\usepackage[margin=1in]{geometry}
\usepackage{hyperref}
\usepackage[numbers,sort&compress]{natbib}
\usepackage{microtype}
\usepackage{xcolor}
\usepackage{booktabs}
\usepackage{enumitem}
\usepackage{listings}
\usepackage{tikz}
\usetikzlibrary{arrows.meta,positioning,shapes.geometric,fit,backgrounds,calc,decorations.pathreplacing}
\usepackage{caption}
\usepackage{subcaption}
\usepackage{mdframed}

\definecolor{teal}{HTML}{1C7293}
\definecolor{teallight}{HTML}{E1F0F2}
\definecolor{coral}{HTML}{B85042}
\definecolor{corallight}{HTML}{F6E7E3}
\definecolor{green}{HTML}{3B6D11}
\definecolor{greenlight}{HTML}{EAF3DE}
\definecolor{sand}{HTML}{F1EFE8}
\definecolor{orange}{HTML}{BA7517}
\definecolor{orangelight}{HTML}{FAEEDA}
\definecolor{darktext}{HTML}{21303A}
\definecolor{mutedtext}{HTML}{5F5E5A}

\newmdenv[
  backgroundcolor=corallight,
  linecolor=coral,
  linewidth=1.2pt,
  roundcorner=4pt,
  innertopmargin=6pt,
  innerbottommargin=6pt,
]{problembox}

\newmdenv[
  backgroundcolor=teallight,
  linecolor=teal,
  linewidth=1.2pt,
  roundcorner=4pt,
  innertopmargin=6pt,
  innerbottommargin=6pt,
]{conceptbox}

\newmdenv[
  backgroundcolor=greenlight,
  linecolor=green,
  linewidth=1.2pt,
  roundcorner=4pt,
  innertopmargin=6pt,
  innerbottommargin=6pt,
]{solutionbox}

\title{\textbf{The Spec Growth Engine}\\
  \large Spec-Anchored, Code-Coupled, Drift-Enforced Architecture\\
  for AI-Assisted Software Development}

\author{
  Hartwig Grabowski\\
  \small Hochschule Offenburg\\
  \small \texttt{hartwig.grabowski@hs-offenburg.de}\\
  \small ORCID: \href{https://orcid.org/0009-0001-4300-2626}{0009-0001-4300-2626}
}

\date{\today}

\begin{document}
\maketitle

\begin{abstract}
AI coding agents dramatically accelerate implementation speed but introduce two
structural failure modes that existing spec-driven approaches do not fully solve:
(1)~\emph{context explosion} -- the agent must reason over an entire repository at
once, degrading output quality as the context window fills; and (2)~\emph{silent
spec-code drift} -- code evolves, the specification does not, and the divergence
becomes invisible until it is costly to repair.
We present the \emph{Spec Growth Engine}, a lightweight framework that addresses
both failure modes through a machine-readable \emph{spec graph} whose nodes carry
explicit contract/design separation, a \emph{Spine} context assembler that scopes
agent context to an ownership path, a \emph{vertical-slice growth protocol} that
enforces hardest-first ordering, and a \emph{drift gate} that makes
spec-code divergence a blocking merge condition.
The design synthesises well-established software engineering principles (Parnas
information hiding, C4, ADRs, Walking Skeleton, Reflexion Models, Fitness
Functions) into a lean, code-coupled, machine-enforced whole -- without the
overhead of heavy-weight frameworks such as RUP or MDA.
\end{abstract}

\section{Introduction}
\label{sec:intro}

Modern AI coding agents can write, refactor, and test code at a pace that far
outstrips traditional developer throughput.  Yet in practice they exhibit two
recurring failure patterns that make the output brittle:

\begin{itemize}
  \item \textbf{Too much at once.}  A single prompt that spans an entire system
        causes the agent to over-engineer, invent structure, or simply ``get lost''
        as the context window fills.  Practitioners call the high-fill regime the
        \emph{Dumb Zone}~\cite{horthy2025}; the underlying degradation with input
        length is documented more rigorously as ``context rot''
        (Section~\ref{sec:context-explosion}).

  \item \textbf{Spec-code divergence.}  The agent changes code; nobody updates
        the specification; the system still passes tests; the divergence is never
        flagged.  Over time, the specification becomes a historical artefact rather
        than a living contract, and future agent runs are guided by a lie.
\end{itemize}

Existing approaches sit at the extremes.  \emph{Spec-first} frameworks
(e.g.\ AWS Kiro~\cite{kiro2025}, Tessl) generate full specifications before any
code, at the cost of upfront overhead and the risk of specifying the wrong thing.
\emph{Spec-as-source} systems (MDA, some model-driven IDEs) generate code from
specifications, but introduce nondeterminism and a fragile single point of truth
that teams consistently reject in practice.

The Spec Growth Engine occupies a deliberately lean middle ground: it is
\emph{spec-anchored} (every node has a spec) yet \emph{code-coupled} (code and
spec evolve together, in the same commit) and \emph{drift-enforced} (divergence
is a blocking merge error, not a discipline problem).

\paragraph{Paper organisation.}
Section~\ref{sec:problems} characterises the two failure modes in depth.
Section~\ref{sec:foundations} reviews the established ideas we build on.
Section~\ref{sec:slices} introduces Vertical Slices and the Spine as our central
scaffolding concepts.
Section~\ref{sec:engine} describes the full Spec Growth Engine.
Section~\ref{sec:workflow} walks through the development workflow.
Section~\ref{sec:discussion} discusses trade-offs.
Section~\ref{sec:conclusion} concludes.

\section{The Two Structural Failure Modes}
\label{sec:problems}

\subsection{Context Explosion: Too Much at Once}
\label{sec:context-explosion}

A coding agent operates on a context bundle -- the set of files and texts it can
attend to during a single generation.  Without scoping, ``give the agent everything
relevant'' quickly means ``give it the whole repository.''

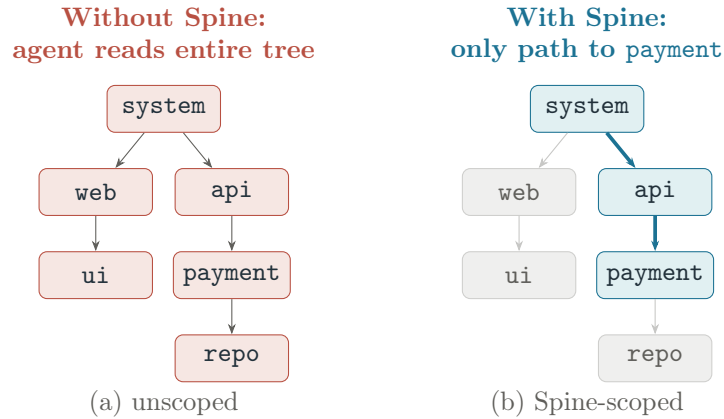
\begin{figure}[h]
\centering
\begin{tikzpicture}[
  every node/.style={font=\small},
  box/.style={draw, rounded corners=3pt, minimum width=1.5cm, minimum height=0.62cm,
              fill=teallight, draw=teal, text=darktext, align=center},
  cross/.style={draw, rounded corners=3pt, minimum width=1.5cm, minimum height=0.62cm,
                fill=corallight, draw=coral, text=darktext, align=center},
  muted/.style={draw, rounded corners=3pt, minimum width=1.5cm, minimum height=0.62cm,
                fill=mutedtext!10, draw=mutedtext!30, text=mutedtext, align=center},
  arrow/.style={-{Stealth[length=4pt]}, draw=mutedtext}
]
  \node[cross] (root)  at (1.3,0)            {\texttt{system}};
  \node[cross] (web)   at (0.4,-1.1)          {\texttt{web}};
  \node[cross] (api)   at (2.2,-1.1)          {\texttt{api}};
  \node[cross] (ui)    at (0.4,-2.2)          {\texttt{ui}};
  \node[cross] (pay)   at (2.2,-2.2)          {\texttt{payment}};
  \node[cross] (repo)  at (2.2,-3.3)          {\texttt{repo}};
  \draw[arrow] (root)--(web); \draw[arrow] (root)--(api);
  \draw[arrow] (web)--(ui);   \draw[arrow] (api)--(pay);
  \draw[arrow] (pay)--(repo);
  \node[above=0.15cm of root,align=center,font=\small\bfseries,text=coral]
       {Without Spine:\\agent reads entire tree};

  \begin{scope}[xshift=5.6cm]
    \node[box]   (root2) at (1.3,0)           {\texttt{system}};
    \node[muted] (web2)  at (0.4,-1.1)         {\texttt{web}};
    \node[box]   (api2)  at (2.2,-1.1)         {\texttt{api}};
    \node[muted] (ui2)   at (0.4,-2.2)         {\texttt{ui}};
    \node[box]   (pay2)  at (2.2,-2.2)         {\texttt{payment}};
    \node[muted] (repo2) at (2.2,-3.3)         {\texttt{repo}};
    \draw[arrow,draw=teal,line width=1.4pt] (root2)--(api2);
    \draw[arrow,draw=teal,line width=1.4pt] (api2)--(pay2);
    \draw[arrow,draw=mutedtext!35] (root2)--(web2);
    \draw[arrow,draw=mutedtext!35] (web2)--(ui2);
    \draw[arrow,draw=mutedtext!35] (pay2)--(repo2);
    \node[above=0.15cm of root2,align=center,font=\small\bfseries,text=teal]
         {With Spine:\\only path to \texttt{payment}};
  \end{scope}

  \node[font=\small,text=mutedtext] at (1.3, -3.9)  {(a) unscoped};
  \node[font=\small,text=mutedtext] at (6.9, -3.9)  {(b) Spine-scoped};
\end{tikzpicture}
\caption{Context scoping via the Spine. Without the Spine (a), an agent working on
  \texttt{payment} reads the entire tree. With the Spine (b), it receives only the
  ownership path from root to \texttt{payment} plus the contracts of declared
  dependencies -- a fraction of the total repository.}
\label{fig:spine-scoping}
\end{figure}

The practical consequence of an unscoped context is twofold:
(1) the agent conflates concerns from unrelated modules; and
(2) it produces ``global fixes'' that touch boundaries it was not asked to change.

A third consequence is that output quality \emph{degrades with context length} --
the effect coding-agent practitioners shorthand as the \emph{Dumb Zone}.  We are
careful to separate the heuristic from the evidence.  The often-quoted figure --
become cautious past roughly 40\% fill and do not treat the last $\sim$60\% as a
working zone~\cite{horthy2025} -- is a \emph{rule of thumb} from coding-agent
practice, not a proven threshold; the exact point depends on model and task.  The
\emph{underlying} effect, however, is well supported by research: models use long
contexts unevenly and retrieve worst from the middle~\cite{liu2023lost}; quality
falls consistently as input grows (``context rot'')~\cite{chroma2025contextrot};
and on long-context coding specifically, performance can collapse at scale --
e.g.\ a strong model dropping from $29\%$ to $3\%$ on a long software-engineering
benchmark as the window grows from $32$K to $256$K
tokens~\cite{longcodebench2025}.  The takeaway is directional, not numerical:
\emph{less, well-chosen context beats more}, which is exactly what the Spine
(Section~\ref{sec:spine}) provides.

The root cause is architectural: there is no principled way to say ``the agent
working on \texttt{payment} is allowed to know \emph{exactly this and nothing
more}.''  Without that boundary, the developer must either give everything
(expensive, noisy) or guess what is relevant (fragile, inconsistent).

\subsection{Code-Spec Divergence: Silent Drift}
\label{sec:drift-problem}

The second failure mode is subtler.  When a specification is maintained as a
separate document, every code change creates a potential divergence between intent
(what the spec claims) and evidence (what the code does).  In practice, developers
update the code and defer the spec update -- permanently.

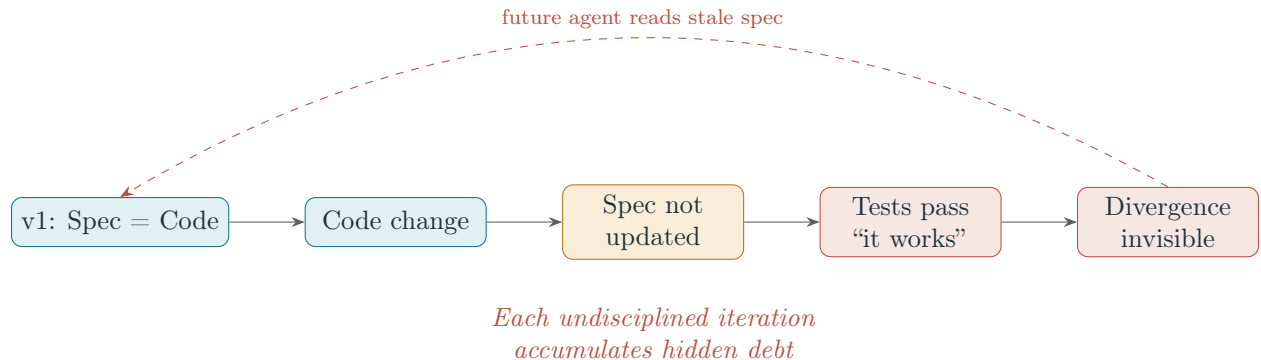
\begin{figure}[h]
\centering
\resizebox{\linewidth}{!}{%
\begin{tikzpicture}[
  node distance=1.0cm and 1.0cm,
  every node/.style={font=\small},
  phase/.style={draw, rounded corners=4pt, minimum width=2.4cm, minimum height=0.65cm,
                fill=teallight, draw=teal, text=darktext, align=center},
  warn/.style={draw, rounded corners=4pt, minimum width=2.4cm, minimum height=0.65cm,
               fill=orangelight, draw=orange, text=darktext, align=center},
  bad/.style={draw, rounded corners=4pt, minimum width=2.4cm, minimum height=0.65cm,
              fill=corallight, draw=coral, text=darktext, align=center},
  arrow/.style={-{Stealth[length=5pt]}, draw=mutedtext},
]
  \node[phase] (s0) {v1: Spec = Code};
  \node[phase, right=of s0] (s1) {Code change};
  \node[warn,  right=of s1] (s2) {Spec not\\updated};
  \node[bad,   right=of s2] (s3) {Tests pass\\``it works''};
  \node[bad,   right=of s3] (s4) {Divergence\\invisible};

  \draw[arrow] (s0)--(s1);
  \draw[arrow] (s1)--(s2);
  \draw[arrow] (s2)--(s3);
  \draw[arrow] (s3)--(s4);

  \node[below=0.5cm of s2, align=center, font=\small\itshape, text=coral]
       {Each undisciplined iteration\\accumulates hidden debt};

  \draw[-{Stealth[length=5pt]},draw=coral,dashed]
       (s4.north) to[bend right=30] node[above,font=\scriptsize,text=coral] {future agent reads stale spec}
       (s0.north);
\end{tikzpicture}}
\caption{The silent drift cycle. A passing test suite does not guarantee that the
  specification and code agree. Each iteration that defers a spec update increases
  the gap; a future coding agent that relies on the stale spec will produce work
  that is inconsistent with actual system behaviour.}
\label{fig:drift-cycle}
\end{figure}

This is not a new observation: Murphy and Notkin identified the same pattern in
their Reflexion Model work~\cite{murphy1995}, and Perry and Wolf coined
``architectural erosion'' in the same era~\cite{perry1992}.  What is new is the
cost: when an AI coding agent generates several hundred lines per minute guided by
a stale spec, the damage from a diverged spec accumulates far faster than in
traditional development.

The failure mode is especially insidious because it remains invisible.  A linter
does not flag it.  CI does not flag it.  The system ships with drift baked in.

\section{Established Foundations}
\label{sec:foundations}

The Spec Growth Engine is not a new paradigm; it is a lean, machine-enforced
synthesis of well-established techniques.  Table~\ref{tab:origins} maps each
mechanism to its intellectual origin.

\begin{table}[h]
\centering
\caption{Origins of the Spec Growth Engine building blocks.}
\label{tab:origins}
\small
\begin{tabular}{@{}ll@{}}
\toprule
\textbf{Mechanism} & \textbf{Established origin} \\
\midrule
Contract vs.\ design visibility & Parnas, \emph{Information Hiding} (1972)~\cite{parnas1972} \\
Module guides per component & Parnas, A-7E project (1979)~\cite{parnas1979} \\
C4 architecture levels & Simon Brown, C4 model~\cite{brown2018} \\
Architecture documentation & ISO/IEC/IEEE~42010, arc42, 4+1 views~\cite{kruchten1995} \\
Architecture Decision Records & Nygard (2011)~\cite{nygard2011} \\
Dependency inversion / acyclic deps & Martin, \emph{Agile Principles}~\cite{martin2002} \\
Vertical slices / Walking Skeleton & Cockburn; Freeman \& Pryce, GOOS~\cite{freeman2009} \\
Hardest/riskiest slice first & Boehm, Spiral model~\cite{boehm1986} \\
Continuous Delivery / DORA & Humble \& Farley~\cite{humble2010}; Forsgren et al.~\cite{forsgren2018} \\
Drift validation & Murphy \& Notkin, Reflexion Models~\cite{murphy1995} \\
Automated fitness functions & Ford, Parsons \& Kua~\cite{ford2022} \\
Component reuse / DRY & McIlroy (1968); Fowler, \emph{Refactoring}~\cite{fowler1999} \\
Bounded Contexts & Evans, DDD (2003)~\cite{evans2003}; Khononov (2021)~\cite{khononov2021} \\
Stage gates / Definition of Done & Cooper~\cite{cooper1990}; Scrum~\cite{schwaber2020} \\
Body of Knowledge reference & SWEBOK v4 (2024)~\cite{swebok2024} \\
\bottomrule
\end{tabular}
\end{table}

Our contribution is the integration layer: a single, lean framework that
enforces all these properties simultaneously via machine-readable artefacts and
blocking gates, at an overhead low enough for daily use with AI agents.

\section{Vertical Slices and the Spine}
\label{sec:slices}

Before describing the full engine, we introduce its two central scaffolding
concepts: \emph{Vertical Slices} (how the system grows) and the \emph{Spine}
(how agent context is bounded).

\subsection{Vertical Slices: Hardest First}
\label{sec:vertical-slices}

A \emph{vertical slice} is a thin, end-to-end-functional increment through the
system -- touching every architectural layer it requires, but only the parts it
requires.  The term originates with the Walking Skeleton pattern~\cite{cockburn2001}
and Tracer Bullets~\cite{hunt1999}: build the thinnest possible path that
demonstrates the system actually works before adding width.

The Spec Growth Engine formalises this into a two-layer rule:

\begin{solutionbox}
\textbf{Layer 1 -- Invariants (up front, deliberate):}
Root invariants, key container boundaries (persistence, security, external
integrations, error taxonomy) are specified before any feature.  They are
\emph{not} just-in-time.  They form the floor below which the architecture cannot
silently erode.

\medskip
\noindent\textbf{Layer 2 -- Features (agile, just-in-time):}
Everything else grows as functional vertical slices, ordered hardest-first.
Slice~\#1 is a tracer bullet through the riskiest integration.  Fakes are
permitted only at the frontier (a later slice's dependency), never on the current
slice's active path.
\end{solutionbox}

This rule prevents the two pathological orderings that both fail:
\emph{breadth-first faking} (hard problems hide in stubs and surface only at the
end) and \emph{pure agile with no floor} (a needed boundary such as persistence
never appears because no feature happened to force it -- the origin of many
hardcoded workarounds in production systems).

A \emph{frontier stub} is not a kind of implementation but a \emph{position}
together with a piece of governance.  Technically it is an ordinary placeholder --
a hollow shell that returns a canned answer and does no real work; the same
artefact on the active path would be a forbidden fake.  What makes it a
\emph{frontier} stub is (i)~\emph{where} it sits -- off the current slice's active
path, standing in for a dependency a later slice will build -- and (ii)~\emph{how}
it is tracked -- as an explicit exception that names its successor slice and keeps
the node at status \texttt{specd}, so it can never be mistaken for done.  Its
purpose is to let the current node declare its dependency \emph{cleanly}: a node
may depend only on a \emph{contract} (Section~\ref{sec:spec-graph}), never on
internal code, so the seam it points at must already expose one.  The frontier
stub supplies exactly that contract -- the interface is fixed now, the
implementation deferred to the slice that owns it.

\begin{figure}[h]
\centering
\begin{tikzpicture}[
  node distance=0.55cm and 0.35cm,
  every node/.style={font=\scriptsize},
  inv/.style={draw, rounded corners=2pt, minimum width=1.9cm, minimum height=0.55cm,
              fill=corallight, draw=coral, text=darktext, align=center},
  feat/.style={draw, rounded corners=2pt, minimum width=1.9cm, minimum height=0.55cm,
               fill=teallight, draw=teal, text=darktext, align=center},
  done/.style={draw, rounded corners=2pt, minimum width=1.9cm, minimum height=0.55cm,
               fill=greenlight, draw=green, text=darktext, align=center},
  fake/.style={draw, rounded corners=2pt, minimum width=1.9cm, minimum height=0.55cm,
               fill=sand, draw=mutedtext, text=mutedtext, align=center,
               dashed},
  arrow/.style={-{Stealth[length=4pt]}, draw=mutedtext!60},
  label/.style={font=\scriptsize\bfseries, align=left, text width=1.7cm, anchor=west},
]
  \node[label, text=coral] (l1hdr) at (0,0) {Layer 1\\Invariants};
  \node[inv, right=0.6cm of l1hdr] (db)   {persistence};
  \node[inv, right=0.3cm of db]    (sec)  {security};
  \node[inv, right=0.3cm of sec]   (ext)  {ext.\ API};

  \node[label, text=teal] (l2hdr) at (0,-1.7) {Layer 2\\Features};
  \node[done, right=0.6cm of l2hdr] (s1) {Slice \#1\\tracer $\checkmark$};
  \node[feat, right=0.3cm of s1]    (s2) {Slice \#2};
  \node[fake, right=0.3cm of s2]    (s3) {Slice \#3\\(frontier)};

  \draw[decorate, decoration={brace,mirror,amplitude=4pt,raise=4pt},
        draw=coral]
       (db.south west) -- (ext.south east)
       node[midway, below=6pt, font=\scriptsize, text=coral]
            {specified before first feature};

  \draw[decorate, decoration={brace,amplitude=4pt,raise=3pt,mirror},
        draw=teal]
       (s1.south west) -- (s3.south east)
       node[midway, below=6pt, font=\scriptsize, text=teal]
            {real end-to-end \quad\quad frontier stub only};
\end{tikzpicture}
\caption{The two-layer growth rule. Layer~1 invariants are specified up front.
  Layer~2 features grow as hardest-first vertical slices.  Frontier stubs are
  permitted only for dependencies that belong to a later slice, never on the
  current slice's active path.}
\label{fig:two-layer}
\end{figure}
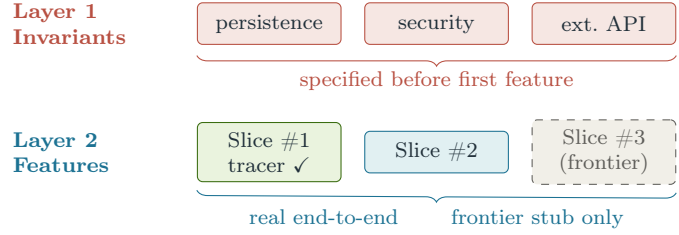

\subsection{The Spine: Scoped Agent Context}
\label{sec:spine}

The \emph{Spine} is the ownership path from the graph root to the current working
node~$N$:
\[
  \text{Spine}(N) \;=\; \text{root} \;\to\; \cdots \;\to\; \text{parent}(N) \;\to\; N
\]

It answers three questions that an agent needs but must not answer by free-form
repository search:
\begin{enumerate}[topsep=2pt,itemsep=2pt]
  \item \emph{Why does $N$ exist?}  The higher-level promise $N$ supports.
  \item \emph{What constraints must $N$ obey?}  Root invariants propagated
        downward.
  \item \emph{What may $N$ use?}  The contracts (not designs) of $N$'s declared
        dependencies -- one hop only.
\end{enumerate}

The context bundle delivered to the coding agent is:
\[
\resizebox{\linewidth}{!}{$\displaystyle
  \text{Context}(N) \;=\;
  \underbrace{\text{RootInvariants}}_{\text{ARCHITECTURE.md}}
  \;+\; \underbrace{\text{SpineContracts}(\text{root} \to N)}_{\text{ownership path}}
  \;+\; N.\text{fullSpec}
  \;+\; \underbrace{\text{Contracts}(\text{deps}(N))}_{\text{one-hop only}}
  \;+\; N.\text{ownCode}
$}
\]

Crucially, the bundle \emph{excludes}: sibling components, dependency designs,
dependency code, transitive dependencies, and ad-hoc grep results.  The assembler
determines what the agent reads; the agent does not search freely.

This is Parnas information hiding~\cite{parnas1972} applied to agent context:
the same boundary that hides a module's design from its neighbours also limits
what the coding agent for that module is allowed to attend to.

\paragraph{ARCHITECTURE.md as a transversal.}
Cross-cutting invariants (principles, ADRs, security policies, naming conventions)
do \emph{not} live in the ownership tree -- they would need to be replicated
everywhere.  Instead, a single \texttt{ARCHITECTURE.md} sits transversally to the
tree and is prepended to every context bundle automatically.  It is not a node;
it has no owner; it ``belongs to'' every node simultaneously.

Concretely, \texttt{ARCHITECTURE.md} \emph{is} the artefact in which the Layer-1
invariants of Section~\ref{sec:vertical-slices} are written down.  The two-layer
rule says \emph{what} must be fixed before any feature (persistence, security,
external integrations, error taxonomy); \texttt{ARCHITECTURE.md} is \emph{where}
those decisions live and the mechanism by which they reach every slice -- because
it is prepended to every Spine, a Layer-1 invariant is, by construction, visible
to every node and enforceable by the drift gate.  Layer~1 is the policy;
\texttt{ARCHITECTURE.md} is its implementation.

The two layers thus map cleanly onto two artefacts.  Layer~2 -- the features --
is realised as the per-node \texttt{SPEC.md} files (Section~\ref{sec:spec-graph}):
each vertical slice adds or updates the contract and design of the nodes along its
path, so the feature layer grows one \texttt{SPEC.md} at a time.  Together the two
are the entire written architecture: one transversal \texttt{ARCHITECTURE.md} for
what holds everywhere, and one \texttt{SPEC.md} per node for what each part
promises.  Nothing else needs to be authored by hand -- the index, the dependency
graph, and the drift report are all \emph{derived} (Section~\ref{sec:engine}).

\section{The Spec Growth Engine}
\label{sec:engine}

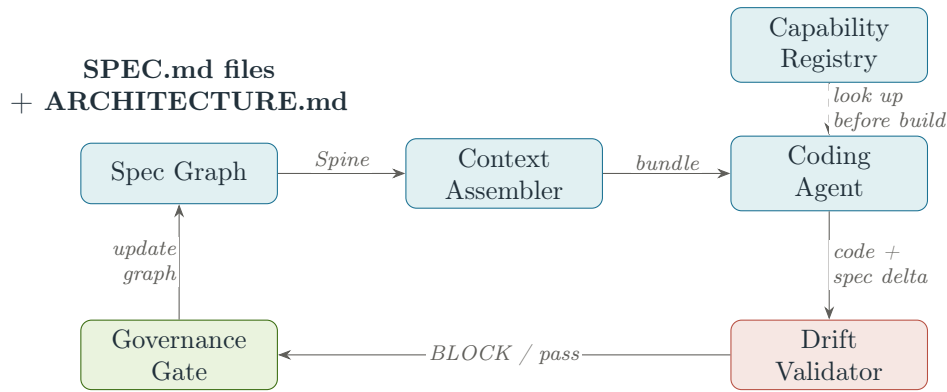
\begin{figure}[h]
\centering
\begin{tikzpicture}[
  every node/.style={font=\small},
  comp/.style={draw, rounded corners=4pt, minimum width=2.6cm, minimum height=0.8cm,
               fill=teallight, draw=teal, text=darktext, align=center},
  cross/.style={draw, rounded corners=4pt, minimum width=2.6cm, minimum height=0.8cm,
                fill=corallight, draw=coral, text=darktext, align=center},
  sol/.style={draw, rounded corners=4pt, minimum width=2.6cm, minimum height=0.8cm,
              fill=greenlight, draw=green, text=darktext, align=center},
  arrow/.style={-{Stealth[length=5pt]}, draw=mutedtext},
  lbl/.style={font=\scriptsize\itshape, text=mutedtext, fill=white, inner sep=1pt},
]
  \node[comp] (sg)    at (0,   0) {Spec Graph};
  \node[comp] (ca)    at (4.3, 0) {Context\\Assembler};
  \node[comp] (agent) at (8.6, 0) {Coding\\Agent};

  \node[comp] (cr) at (8.6, 1.7) {Capability\\Registry};

  \node[cross] (dv) at (8.6, -2.4) {Drift\\Validator};
  \node[sol]   (gg) at (0,   -2.4) {Governance\\Gate};

  \draw[arrow] (sg) -- node[above,lbl]{Spine}  (ca);
  \draw[arrow] (ca) -- node[above,lbl]{bundle} (agent);

  \draw[arrow,dashed] (cr.south) -- node[right,lbl,align=left]{look up\\before build} (agent.north);

  \draw[arrow] (agent.south) -- node[right,lbl,align=left]{code +\\spec delta} (dv.north);
  \draw[arrow] (dv.west) -- node[lbl]{BLOCK / pass} (gg.east);
  \draw[arrow] (gg.north) -- node[left,lbl,align=right]{update\\graph} (sg.south);

  \node[above=0.25cm of sg, align=center, font=\small\bfseries, text=darktext]
       {SPEC.md files\\+ ARCHITECTURE.md};
\end{tikzpicture}
\caption{High-level architecture of the Spec Growth Engine.  The Spec Graph feeds
  the Context Assembler, which scopes what the Coding Agent reads.  The agent's
  output (code + spec delta) passes through the Drift Validator and then the
  Governance Gate before updating the graph.  The Capability Registry provides a
  horizontal reuse axis.}
\label{fig:engine-overview}
\end{figure}

The engine consists of five interlocking components:
the \emph{Spec Graph}, the \emph{Context Assembler}, the \emph{Drift Validator},
the \emph{Governance Gates}, and the \emph{Capability Registry}.
Figure~\ref{fig:engine-overview} shows how they interact.
These components are operated by four actors with distinct authority
(Section~\ref{sec:roles}) and mutate the graph through a fixed set of growth
rules (Section~\ref{sec:growth-rules}).

\subsection{Actors and Responsibilities}
\label{sec:roles}

Four actors operate the engine, separated by the authority each carries.  The
separation is the same Parnas boundary applied to people and agents: who is
allowed to change the outward surface (high blast radius) versus the inward
surface (local, reversible).

\begin{table}[h]
\centering
\caption{The four actors and their authority.}
\label{tab:roles}
\small
\begin{tabular}{@{}clp{9.2cm}@{}}
\toprule
& \textbf{Actor} & \textbf{Authority / responsibility} \\
\midrule
\textbf{H} & Human Architect & Decision rights over boundaries, public contracts,
                               and root invariants; approves consequential
                               changes; grants exceptions. \\
\textbf{P} & Planner Agent   & Proposes decomposition; drafts contracts,
                               acceptance criteria, and design. Authors intent,
                               subject to a gate. \\
\textbf{C} & Coding Agent    & Implements \emph{one} node inside its context
                               bundle; maintains that node's design and declares
                               its observed dependencies. \\
\textbf{E} & Engine          & Deterministic: derives the index, validates drift,
                               enforces gates, proposes diffs. Authors
                               \emph{no} intent. \\
\bottomrule
\end{tabular}
\end{table}

The critical line is between the Engine and the agents: the Engine is purely
mechanical and never invents architecture, while the Planner and Coding agents
author intent but only the Human Architect may approve changes to the outward
surface.  This is what makes ``the AI updates the spec in the same commit''
(Section~\ref{sec:drift}) safe -- the agent proposes the spec delta, the Engine
checks it against the evidence, and the Human approves only the contract-level
lines.

\subsection{The Spec Graph: Nodes and Edges}
\label{sec:spec-graph}

Every architectural entity is a \emph{node} in a graph, materialised as exactly
one \texttt{SPEC.md} file whose \texttt{kind} field declares its level (source
files belong to their node and carry at most a short header).  Nodes are organised
at four C4~\cite{brown2018} zoom levels:

\[
  \texttt{system} \;\to\; \texttt{container} \;\to\; \texttt{component} \;\to\; \texttt{code}
\]

Each node has two orthogonal views:

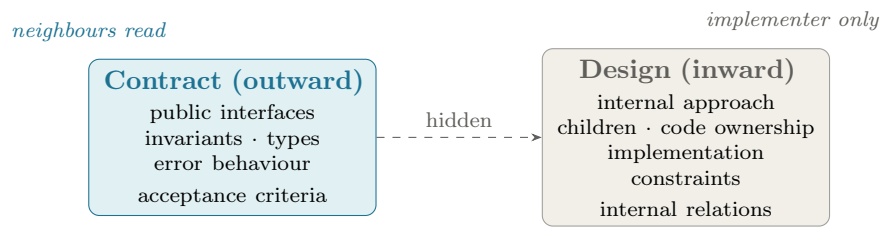
\begin{center}
\begin{tikzpicture}[
  every node/.style={font=\small},
  half/.style={draw, rounded corners=4pt, minimum width=3.8cm, minimum height=1.6cm,
               align=center, text width=3.4cm},
]
  \node[half, fill=teallight, draw=teal] (con) at (0,0) {
    \textbf{\textcolor{teal}{Contract (outward)}}\\[2pt]
    \scriptsize public interfaces\\
    invariants · types\\
    error behaviour\\
    acceptance criteria
  };
  \node[half, fill=sand, draw=mutedtext!60] (des) at (6.0,0) {
    \textbf{\textcolor{mutedtext}{Design (inward)}}\\[2pt]
    \scriptsize internal approach\\
    children · code ownership\\
    implementation constraints\\
    internal relations
  };
  \node[above=0.1cm of con.north west,font=\scriptsize\itshape,text=teal]
       {neighbours read};
  \node[above=0.1cm of des.north east,font=\scriptsize\itshape,text=mutedtext,xshift=-0.5cm]
       {implementer only};
  \draw[-{Stealth[length=4pt]},draw=mutedtext,dashed] (con.east)--(des.west)
       node[midway,above,font=\scriptsize,text=mutedtext]{hidden};
\end{tikzpicture}
\end{center}

This is Parnas information hiding~\cite{parnas1972} applied at every level of the
C4 hierarchy: neighbours are permitted to read only the contract; the design is
visible only to the node's implementer and its parent.

There are two distinct edge types:

\begin{itemize}
  \item \textbf{Ownership edges} form a tree.  Each node has exactly one parent.
        The semantics are ``consists of / realised by'': a parent's contract is
        the sum of its children's contracts.  Ownership edges define the Spine.

  \item \textbf{Dependency edges} form a DAG.  The semantics are ``N may use
        D's contract, never D's design or code.''  Dependencies are acyclic at
        the architecture level.  Context is one hop only.
\end{itemize}

\subsection{The Context Assembler}
\label{sec:assembler}

The context assembler is a deterministic function:
\[
  \text{assemble}(N, G) \;\to\; \text{bundle}
\]
where $G$ is the spec graph.  It does not search; it traverses the graph along
the Spine and declared dependency edges.  If the bundle is insufficient for a
task, the assembler reports a structured diagnostic -- missing contract, undeclared
dependency, missing acceptance criterion -- rather than silently falling back to
free search.  The default repair is to fix the spec graph, not to widen the
context.

\subsection{Drift Validation: Intent Graph vs.\ Evidence Graph}
\label{sec:drift}

Drift validation compares two derived graphs:

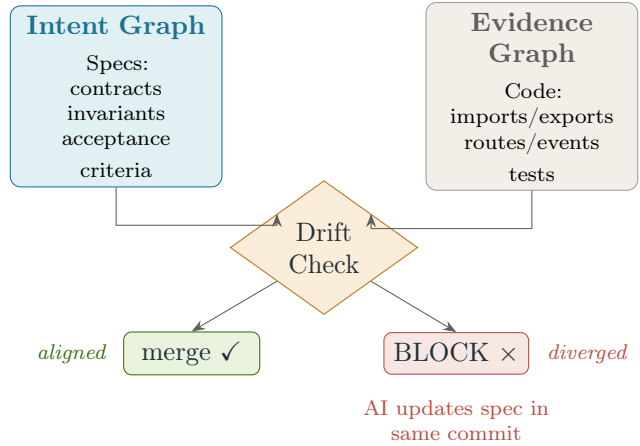
\begin{figure}[h]
\centering
\begin{tikzpicture}[
  every node/.style={font=\small},
  box/.style={draw, rounded corners=4pt, minimum width=2.8cm, minimum height=2.2cm,
              text width=2.4cm, align=center},
  gate/.style={draw, diamond, aspect=1.4, minimum width=2.0cm, minimum height=1.1cm,
               fill=orangelight, draw=orange, text=darktext, align=center, font=\small},
  ok/.style={draw, rounded corners=3pt, minimum width=1.8cm, minimum height=0.55cm,
             fill=greenlight, draw=green, text=darktext, align=center},
  fail/.style={draw, rounded corners=3pt, minimum width=1.8cm, minimum height=0.55cm,
               fill=corallight, draw=coral, text=darktext, align=center},
  arrow/.style={-{Stealth[length=5pt]}, draw=mutedtext},
]
  \node[box, fill=teallight, draw=teal] (intent) at (0,0) {
    \textbf{\textcolor{teal}{Intent Graph}}\\[4pt]
    \scriptsize Specs:\\
    contracts\\
    invariants\\
    acceptance\\criteria
  };
  \node[box, fill=sand, draw=mutedtext!60] (evid) at (5.5,0) {
    \textbf{\textcolor{mutedtext}{Evidence Graph}}\\[4pt]
    \scriptsize Code:\\
    imports/exports\\
    routes/events\\
    tests
  };
  \node[gate] (check) at (2.75,-2.0) {Drift\\Check};
  \node[ok]   (pass)  at (1.0,-3.4) {merge $\checkmark$};
  \node[fail] (block) at (4.5,-3.4) {BLOCK \texttimes};

  \draw[arrow] (intent.south) -- ++(0,-0.5) -| (check.north west);
  \draw[arrow] (evid.south)   -- ++(0,-0.5) -| (check.north east);
  \draw[arrow] (check.south west) -- (pass.north);
  \draw[arrow] (check.south east) -- (block.north);

  \node[left=0.1cm of pass, font=\scriptsize\itshape, text=green]   {aligned};
  \node[right=0.1cm of block, font=\scriptsize\itshape, text=coral] {diverged};

  \node[below=0.2cm of block, font=\scriptsize, text=coral, align=center]
       {AI updates spec in\\same commit};
\end{tikzpicture}
\caption{The drift validation gate.  The engine derives the Intent Graph from
  SPEC.md files and the Evidence Graph from static code analysis.  A commit
  where the two graphs disagree is blocked.  The AI agent updates the affected
  spec in the same commit; the human approves only contract-level changes.}
\label{fig:drift-gate}
\end{figure}

Hard errors that block merge unconditionally:
\begin{itemize}[noitemsep]
  \item Orphan code (a source file with no spec owner)
  \item Undeclared dependency (code imports across a spec boundary without a
        declared edge)
  \item Dependency bypasses contract (code imports internal files of another node)
  \item Missing dependency contract (a target node has no contract)
\end{itemize}

Warnings that are reported but do not block:
\begin{itemize}[noitemsep]
  \item Declared dependency with no code evidence
  \item Public export not mentioned in the contract
  \item Contract behaviour without test evidence
\end{itemize}

The key invariant: spec and code may \emph{never diverge silently}.  Keeping
them in sync is not left to discipline -- the AI agent updates the affected spec
in the same commit, and the human reviews only contract-level changes.  This
transforms drift from a social/process problem into a structural impossibility.

\subsection{Governance Gates}
\label{sec:governance}

Not every change carries the same risk.  The governance model follows a single
principle:

\begin{conceptbox}
\emph{Authority follows blast radius and reversibility.}
Changes to the outward surface (root invariants, structure, public contract)
have high blast radius and low reversibility, so they require a gate.
Changes to the inward surface (internal design, owned code) are local and
reversible, so they may be made autonomously.
\end{conceptbox}

This yields three gate levels:
\begin{itemize}
  \item \textbf{HARD} -- human approval required before merge (root invariant
        change, new container boundary, breaking contract change).
  \item \textbf{SOFT} -- agent may proceed; human review requested
        asynchronously (new component boundary, additive contract change).
  \item \textbf{AUTO} -- no human; engine policy decides (internal design,
        pure refactoring within an owned node).
\end{itemize}

This is the contract/design visibility rule re-expressed as decision rights.

\subsection{Capability Registry}
\label{sec:registry}

The Capability Registry is the horizontal reuse axis of the engine.  When an
agent is about to implement a new capability, it queries the registry first:

\begin{lstlisting}
spec query cap:pdf-to-text
# -> found: shared/pdf-parser (owner: infra.pdf)
# -> use contract, do not rebuild
\end{lstlisting}

A capability is identified by a stable id (\texttt{cap:<name>}) and has exactly
one providing node.  Two nodes that provide the same capability are flagged as a
duplicate -- the ``DRY at architecture level'' check.

This prevents the failure mode documented in practice: two teams build the same
PDF reader in parallel; both ship; neither is identified as a duplicate until a
third team needs it and finds two incompatible implementations.

\subsection{The Growth Rules}
\label{sec:growth-rules}

The graph is never edited freely.  Every change is classified and triggers a
fixed rule that names exactly which artefacts must move together.  This is what
keeps growth incremental and contract-preserving rather than a free-form rewrite.

\begin{table}[h]
\centering
\caption{The six growth rules. Each rule fires on a change classification and
  prescribes the minimal, coupled set of artefacts to update.}
\label{tab:growth-rules}
\small
\begin{tabular}{@{}p{2.6cm}p{4.1cm}p{6.4cm}@{}}
\toprule
\textbf{Rule} & \textbf{Fires when} & \textbf{Required actions} \\
\midrule
\textbf{1. Behaviour} &
  a change adds observable behaviour &
  update the affected node's contract; add/update acceptance criteria;
  add/update code and tests. \\
\textbf{2. Decomposition} &
  a node grows internally structured or too large &
  add the new child to the parent's design; create a minimal child spec if it is
  a real boundary; add the ownership edge; map the child to its code. \\
\textbf{3. Dependency} &
  node $N$ starts using node $D$ &
  declare $N \to D$; ensure $D$ exposes a contract; allow $N$ to read only $D$'s
  contract; forbid reading $D$'s design or code. \\
\textbf{4. Boundary} &
  a public promise changes &
  update the contract and acceptance criteria; find direct consumers; run or add
  contract/consumer tests. \\
\textbf{5. Internal design} &
  only the internal implementation changes &
  update the node's design only if the change is durable; keep the contract
  unchanged; do not touch neighbour specs. \\
\textbf{6. C4 promotion} &
  the change affects a higher C4 level &
  update the highest affected node; add only the minimal new contract/design the
  current change needs; do not pre-specify unrelated future components. \\
\bottomrule
\end{tabular}
\end{table}

Two properties follow directly from this rule set.  First, decomposition (Rule~2)
is \emph{contract-preserving}: because the children's contracts sum to the
parent's, splitting a node into children does not change its outward promise --
neighbours are unaffected.  Second, the rules are deliberately
\emph{minimal-touch}: an internal-only change (Rule~5) must \emph{not} ripple into
neighbour specs, which is what bounds the blast radius and keeps most changes in
the AUTO gate.  The rules never instruct ``read more files''; the default repair
for an insufficient context is to fix the spec graph, not to widen it.

\section{Development Workflow}
\label{sec:workflow}

The full workflow for a feature is deterministic and tool-supported:

\begin{enumerate}[leftmargin=*, topsep=4pt, itemsep=3pt]
  \item \textbf{Select} the story or change; identify the target node~$N$ (or
        create it via a Planner Agent proposal with SOFT gate).

  \item \textbf{Locate} $N$ in the spec graph; run the growth rules
        (Table~\ref{tab:growth-rules}) to update the minimal affected specs.

  \item \textbf{Assemble} $\text{Context}(N)$ deterministically; the agent
        receives exactly this bundle.

  \item \textbf{Implement} -- the Coding Agent works within the bundle; its
        output is code + a spec delta (the same-commit contract update).

  \item \textbf{Validate} -- the drift validator compares Intent Graph and
        Evidence Graph; hard errors block; warnings are reported.

  \item \textbf{Gate} -- the governance gate checks blast radius; HARD changes
        go to human review; AUTO changes land without review.

  \item \textbf{Update} the spec graph (the index is regenerated, not
        hand-maintained).
\end{enumerate}

The graph grows with working software.  The invariant at each step is:
\begin{center}
  \emph{spec and code are always aligned, or the commit does not land.}
\end{center}

\section{Worked Example: Growing a Checkout}
\label{sec:example}

To make the machinery concrete, we trace a small e-commerce system from
``Hello World'' to a checkout with discount codes -- the same domain used in the
figures throughout.

\paragraph{The failure we are avoiding.}
Consider first what happens \emph{without} an architectural floor.  A
curated-product finder is grown purely agile: feature after feature is added, and
because no single feature ever \emph{forces} persistence, the product list ends up
as a 312-line hardcoded array instead of a database.  Nothing was wrong at any
single step; the boundary simply never appeared.  This is the silent
under-architecture of Section~\ref{sec:vertical-slices}.  The two-layer rule
prevents exactly this: persistence is a Layer-1 invariant, so slice~\#1 already
runs through a real store and the database \emph{cannot} be forgotten.

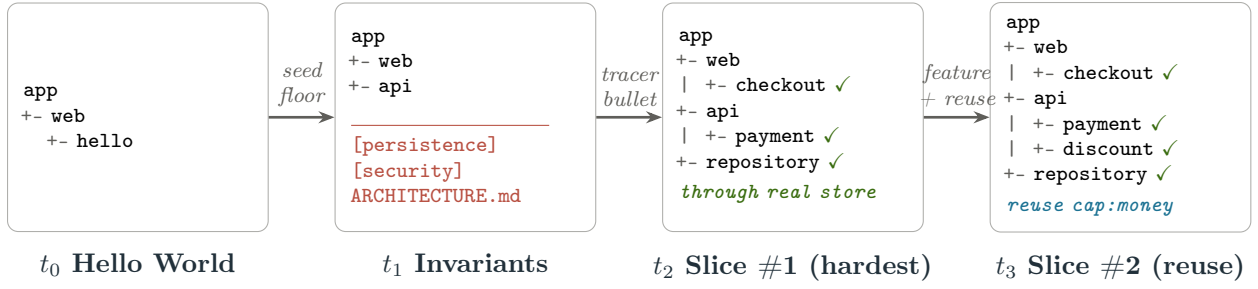
\begin{figure}[h]
\centering
\resizebox{\linewidth}{!}{%
\begin{tikzpicture}[
  every node/.style={font=\small},
  panel/.style={draw, rounded corners=4pt, align=left, text width=3.0cm,
                inner sep=6pt, minimum height=3.0cm, fill=white, draw=mutedtext!50,
                font=\ttfamily\scriptsize},
  pcap/.style={font=\small\bfseries, text=darktext, align=center},
  arrow/.style={-{Stealth[length=6pt]}, draw=mutedtext, line width=1pt},
  albl/.style={font=\scriptsize\itshape, text=mutedtext, align=center},
]
  \node[panel] (p0) at (0,0) {
    app\\
    \textcolor{mutedtext}{+-}~web\\
    \textcolor{mutedtext}{~~~+-}~hello
  };
  \node[panel] (p1) at (4.3,0) {
    app\\
    \textcolor{mutedtext}{+-}~web\\
    \textcolor{mutedtext}{+-}~api\\[3pt]
    \textcolor{coral}{\rule{2.6cm}{0.4pt}}\\
    \textcolor{coral}{[persistence]}\\
    \textcolor{coral}{[security]}\\
    \textcolor{coral}{ARCHITECTURE.md}
  };
  \node[panel] (p2) at (8.6,0) {
    app\\
    \textcolor{mutedtext}{+-}~web\\
    \textcolor{mutedtext}{|~~+-}~checkout~\textcolor{green}{$\checkmark$}\\
    \textcolor{mutedtext}{+-}~api\\
    \textcolor{mutedtext}{|~~+-}~payment~\textcolor{green}{$\checkmark$}\\
    \textcolor{mutedtext}{+-}~repository~\textcolor{green}{$\checkmark$}\\[2pt]
    \textcolor{green}{\itshape through real store}
  };
  \node[panel] (p3) at (12.9,0) {
    app\\
    \textcolor{mutedtext}{+-}~web\\
    \textcolor{mutedtext}{|~~+-}~checkout~\textcolor{green}{$\checkmark$}\\
    \textcolor{mutedtext}{+-}~api\\
    \textcolor{mutedtext}{|~~+-}~payment~\textcolor{green}{$\checkmark$}\\
    \textcolor{mutedtext}{|~~+-}~discount~\textcolor{green}{$\checkmark$}\\
    \textcolor{mutedtext}{+-}~repository~\textcolor{green}{$\checkmark$}\\[2pt]
    \textcolor{teal}{\itshape reuse cap:money}
  };

  \node[pcap, below=0.15cm of p0] {$t_0$ Hello World};
  \node[pcap, below=0.15cm of p1] {$t_1$ Invariants};
  \node[pcap, below=0.15cm of p2] {$t_2$ Slice \#1 (hardest)};
  \node[pcap, below=0.15cm of p3] {$t_3$ Slice \#2 (reuse)};

  \draw[arrow] (p0.east) -- (p1.west) node[midway,above,albl]{seed\\floor};
  \draw[arrow] (p1.east) -- (p2.west) node[midway,above,albl]{tracer\\bullet};
  \draw[arrow] (p2.east) -- (p3.west) node[midway,above,albl]{feature\\+ reuse};
\end{tikzpicture}}
\caption{The spec graph growing with working software.  Layer-1 invariants
  (coral) are seeded before any feature; each subsequent slice is implemented for
  real end-to-end (green check).  The database appears at $t_1$ as an invariant,
  not by accident -- so it can never be ``forgotten''.}
\label{fig:growth-strip}
\end{figure}

\paragraph{How the graph grows.}
Figure~\ref{fig:growth-strip} shows four snapshots.  At $t_0$ the system is a
trivial skeleton.  At $t_1$ the Layer-1 invariants are seeded deliberately
(persistence, security) into \texttt{ARCHITECTURE.md} -- before any feature.  At
$t_2$ the hardest-first slice (\texttt{checkout} $\to$ \texttt{payment} $\to$
real \texttt{repository}) is implemented end-to-end for real.  At $t_3$ a second
slice adds \texttt{discount}, which reuses the shared \texttt{money} capability
rather than rebuilding it.

\paragraph{One slice through the machinery.}
Table~\ref{tab:slice-trace} traces the $t_2 \to t_3$ slice -- ``add a discount
code to checkout'' -- through the full workflow of
Section~\ref{sec:workflow}, naming the acting role, the growth rules that fire,
and the gate decision.  It shows the five components and four actors working
together on a single concrete change.

\begin{table}[h]
\centering
\caption{Tracing the slice ``add discount code to checkout'' ($t_2 \to t_3$)
  through the workflow. Roles: H/P/C/E (Table~\ref{tab:roles});
  rules from Table~\ref{tab:growth-rules}.}
\label{tab:slice-trace}
\small
\begin{tabular}{@{}p{1.9cm}cp{9.3cm}@{}}
\toprule
\textbf{Step} & \textbf{Role} & \textbf{Concrete action} \\
\midrule
Select &
  P\,$\to$\,H &
  Planner proposes a new component \texttt{api.discount}; a new component
  boundary is a SOFT gate. \\
Grow &
  P/E &
  Rule~1 (Behaviour: the checkout total changes) and Rule~3 (Dependency:
  \texttt{checkout}\,$\to$\,\texttt{discount}) fire; contract and acceptance
  criteria are drafted. \\
Assemble &
  E &
  $\text{Context} = $ root invariants $+$ Spine(\texttt{app}\,$\triangleright$\,%
  \texttt{api}\,$\triangleright$\,\texttt{checkout}) $+$ \texttt{discount} spec
  $+$ Contract(\texttt{cap:money}) $+$ own code. No sibling, no dependency code. \\
Implement &
  C &
  Coding agent writes the discount logic and tests \emph{within} the bundle, and
  emits the spec delta in the same commit. \\
Validate &
  E &
  Evidence edges \texttt{checkout}\,$\to$\,\texttt{discount} and
  \texttt{checkout}\,$\to$\,\texttt{money} match the declared dependencies; no
  orphan code; acceptance criteria have test evidence $\Rightarrow$ pass. \\
Gate &
  H &
  The contract change is additive $\Rightarrow$ SOFT: auto-approved, human
  notified asynchronously. \\
Update &
  E &
  Index regenerated; the graph is now at snapshot $t_3$
  (Figure~\ref{fig:growth-strip}). \\
\bottomrule
\end{tabular}
\end{table}

Had the agent instead imported \texttt{discount} \emph{internals} from
\texttt{checkout}, or shipped code with no matching declared edge, the Validate
step would have produced a hard error and the commit would not have landed --
the no-silent-drift invariant of Section~\ref{sec:drift} in action.

\section{Related Work}
\label{sec:related}

The contemporary tools for spec-driven development are best surveyed by
B\"ockeler~\cite{boeckeler2025}, who identifies a ``semantic diffusion'' in the
term and separates it into three maturity levels that we adopt as our axis of
comparison: \emph{spec-first} (specs guide an initial generation, then are
discarded), \emph{spec-anchored} (specs are living artefacts kept in sync with
code), and \emph{spec-as-source} (specs are the single source of truth from which
code is generated).  Table~\ref{tab:related} positions the major tools and the
Spec Growth Engine on this axis.

\begin{table}[h]
\centering
\caption{Spec-driven tools positioned on B\"ockeler's
  maturity axis~\cite{boeckeler2025}.}
\label{tab:related}
\small
\begin{tabular}{@{}lll@{}}
\toprule
\textbf{Tool} & \textbf{Position} & \textbf{Spec--code relationship} \\
\midrule
Kiro~\cite{kiro2025}        & spec-first    & specs deleted after the feature ships \\
Spec Kit~\cite{speckit2025} & spec-first\textsuperscript{*} & aspires to anchored; branch-per-change in practice \\
Tessl~\cite{tessl2025}      & anchored / as-source & bidirectional sync; 1:1 spec-to-file mapping \\
\textbf{Spec Growth Engine} & \textbf{anchored, code-coupled} & \textbf{drift gate; code primary, spec verified} \\
\bottomrule
\end{tabular}\\[2pt]
{\footnotesize \textsuperscript{*}Aspires to spec-anchored; behaves spec-first in practice~\cite{boeckeler2025}.}
\end{table}

\paragraph{Kiro.}
Amazon's Kiro~\cite{kiro2025} structures work as three markdown documents
(Requirements $\to$ Design $\to$ Tasks).  It is squarely spec-first: the specs
guide generation but are not maintained afterward -- they are deleted once the
feature is complete, and a fresh set is created for the next change.  B\"ockeler
notes it is lightweight and easy to review, but disproportionately verbose for
small problems (turning a minor bug into sixteen acceptance criteria).  The Spec
Growth Engine shares Kiro's preference for human-reviewable markdown but rejects
the throw-away model: a node's \texttt{SPEC.md} persists and is the artefact the
drift gate checks against.

\paragraph{GitHub Spec Kit.}
Spec Kit~\cite{speckit2025} adds a project ``constitution'' and a
Constitution $\to$ Specify $\to$ Plan $\to$ Tasks loop driven by slash commands.
It \emph{aspires} to spec-anchored living artefacts, but in practice it behaves
spec-first, generating roughly eight files per specification with extensive
checklists and creating a branch per change request~\cite{boeckeler2025}.  Two of
its reported weaknesses directly motivate our design choices: the volume of
generated markdown, and agents frequently ignoring detailed instructions.  The
Spec Growth Engine attacks the first with a strict one-\texttt{SPEC.md}-per-node
rule and a generated (not hand-written) index, and the second by not relying on
the agent's good behaviour at all -- conformance is enforced by the drift gate
rather than requested in prose.  Spec Kit's ``constitution'' is the closest
analogue to our transversal \texttt{ARCHITECTURE.md}.

\paragraph{Tessl.}
Tessl~\cite{tessl2025} is the only tool explicitly pursuing the spec-anchored and
spec-as-source levels.  It marks generated code with comments, maintains
bidirectional sync, and aims for a 1:1 spec-to-file mapping with low abstraction
per file (reducing the interpretation steps an LLM must make).  B\"ockeler's key
observation is that nondeterminism persists despite precise specs, mirroring the
historical difficulties of model-driven development (MDA).  This is precisely the
line the Spec Growth Engine refuses to cross: we are spec-anchored \emph{like}
Tessl's aspiration, but we keep \emph{code as primary} and treat the spec as a
verified contract -- never as the generative source.  Where Tessl moves toward
spec-as-source and inherits MDA's nondeterminism, we hold at spec-anchored and
make alignment a property the drift gate \emph{checks}, not one a generator must
\emph{produce}.

\paragraph{Positioning.}
Across the three tools, the recurring failure is at the two extremes B\"ockeler
identifies: spec-first tools let the spec go stale (Kiro deletes it; Spec Kit
branches away from it), while spec-as-source tools (Tessl's frontier, MDA) buy
synchronisation at the cost of nondeterminism and a model the developer must
program instead of code.  The Spec Growth Engine occupies the deliberately narrow
middle -- spec-anchored and code-coupled -- and makes that position tenable by
enforcing alignment with a blocking gate rather than either abandoning the spec
or regenerating the code.

\section{Discussion}
\label{sec:discussion}

\paragraph{Overhead.}
The additional artefact is one \texttt{SPEC.md} per architectural node
(component or above) and one \texttt{ARCHITECTURE.md} per system.  Source files
carry at most a short front-matter header for non-obvious roles.  This is
substantially less than traditional architecture documentation, and the overhead
is partially offset by the context savings (a Spine-scoped bundle is far smaller
than a whole-repository context).

\paragraph{Limitations.}
The current design requires a static dependency graph derivable from source files
(imports, routes, events).  Highly dynamic systems (dependency injection at
runtime, plugin architectures) require explicit annotation of late-bound
dependencies.  The governance gate overhead is real for fast-moving teams; the
SOFT gate mitigates this but does not eliminate it.

\paragraph{Relation to SWEBOK v4.}
SWEBOK~v4~\cite{swebok2024} organises the body of knowledge across Knowledge Areas
including Software Architecture, Software Design, Software Testing, Software
Construction, and the new Software Engineering Operations KA.  The Spec Growth
Engine draws on all of these: the spec graph corresponds to Architecture and
Design KAs; drift validation to Testing and Quality; the governance model to
SE~Management; the vertical slice protocol to the Process KA.  We regard SWEBOK~v4
as the background frame of reference against which the engine's positioning can be
checked.

\section{Conclusion}
\label{sec:conclusion}

We presented the Spec Growth Engine, a lean framework for AI-assisted software
development that addresses context explosion and silent spec-code drift through
five interlocking mechanisms: a structured spec graph (C4 levels, contract/design
separation), a Spine-based context assembler, a vertical-slice growth protocol
(hardest-first, two-layer rule), a blocking drift gate, and a governance model
that matches review overhead to change blast radius.

The framework synthesises well-established software engineering principles without
introducing a new paradigm.  Its distinguishing characteristic is machine
enforcement: the properties that classical approaches left to discipline (Parnas
information hiding, no architectural erosion, DRY at architecture level) are here
encoded as blocking gates and deterministic derivations.

The result is a system that is spec-anchored but code-coupled, agile enough to
grow one slice at a time, and strict enough to prevent the hidden coupling that
makes AI-assisted codebases brittle over time.

\paragraph{Availability.}
The method documents and an implementation of the Spec Growth Engine -- engine,
command-line interface and cockpit -- are available under the Apache License~2.0
at\\
\url{https://github.com/grabow/spec-growth-engine}.\\
The implementation, and how agents grow the spec on top of it, is described in a
companion paper~\cite{grabowski2026impl}.

\bibliographystyle{plainnat}
\bibliography{refs}

\end{document}